\documentclass[sigconf]{acmart}
\setcopyright{none}
\renewcommand\footnotetextcopyrightpermission[1]{}
\usepackage{graphicx}
\usepackage{amsmath}
\usepackage{microtype}
\usepackage{url}
\providecommand{\tightlist}{\setlength{\itemsep}{0pt}\setlength{\parskip}{0pt}}
\begin{document}
\title{Incentives and Market Structure in Intent-Based Exchanges: Evidence from a Solver-Reward Reform}
\author{Ruiyang Zhang}
\affiliation{\institution{Ryonix Labs Inc. \& Flock.io}\country{}}
\email{zhangruiyang36@gmail.com}
\begin{abstract}
Intent-based decentralized exchanges delegate execution to a competitive
class of agents --- \emph{solvers} --- whose behavior is shaped by
protocol-designed reward rules. We measure how a change to those rules
reshapes \emph{who captures value}, and we test the mechanism, using a
governance-dated natural experiment: CoW Protocol's CIP-74 (effective
8 December 2025), which replaced a fixed solver-reward cap with one tied
to protocol revenue and introduced an ad-valorem volume fee. Using daily
solver shares over 395 days, we find that the reform reallocated trading
\emph{value} by order size. The robust signature is a monotone size
gradient: concentration \emph{fell} in small orders and \emph{rose} in
large ones, monotonically across four order-value buckets (volume-HHI
level break -0.086, -0.052, +0.025, +0.085 from smallest to largest;
Spearman $\rho$=1.00, exact permutation p=0.042) --- a pattern that
survives dropping the single largest solver. Aggregate concentration also
rose (volume-weighted HHI 0.176$\rightarrow$0.241; top-solver share
0.296$\rightarrow$0.395), but we show this aggregate movement is
substantially carried by the incumbent top solver and report it with that
caveat; the \emph{order-size reallocation} is the result that survives
leave-one-solver-out. The effect is a \emph{value}, not a trade-count,
phenomenon: by number of trades the market \emph{de}-concentrated
(count-HHI -0.060). A simple solver-economics model rationalizes the
pattern --- an ad-valorem fee is competitively neutral, while a
revenue-linked reward cap raises the marginal payoff to inventory-rich
solvers on high-value orders --- the empirical counterpart to theoretical
predictions of restricted entry (Chitra et al.~2024). The timing is
specific to CoW: a control venue (UniswapX) has no sharp concentration
break aligned to the reform date. We detect no change in average
execution quality (the design bounds any change below $\approx$7 bps). A
triple-difference exploiting the February 2026 fee cut on stable pairs is
directionally consistent with the reward channel ($\delta$=-0.042,
predicted sign) but underpowered to confirm it --- a \emph{bounded null}.
Reward design measurably reallocates who captures value in intent
markets, without moving the average price users receive.
\end{abstract}
\maketitle

\section{1. Introduction}

Intent-based exchanges (UniswapX, CoW Protocol, 1inch Fusion, and the
cross-chain ERC-7683 ecosystem) have become a dominant venue for
on-chain trading. Rather than specifying an execution path, a user signs
a desired outcome and a competitive set of \emph{solvers} bids to
fulfill it; the winning solver sources liquidity, often from its own
inventory, and is compensated through a protocol-defined reward
mechanism. Because solvers are the agents who actually set prices and
capture surplus, the rules that pay them are not a back-office detail
--- they are the central lever shaping competition, and ultimately the
price users receive.

Yet the causal effect of solver-reward design on market outcomes is
essentially unstudied. Prior empirical work measures \emph{user-side}
execution quality across venues (Yuminaga et al.~2025; Bachu, Wan \&
Moallemi 2024); theoretical work models solver entry and competition
(Chitra et al.~2024); and practitioner discussion (CoW governance
forums) debates reward effects descriptively. None identifies the causal
impact of an actual reward-rule change, with controls and inference.

We provide the first such estimate, exploiting a clean natural
experiment. On 8 December 2025, CoW Protocol enacted CIP-74, which (i)
replaced a fixed per-auction reward cap with a cap tied to the protocol
revenue of the winning solution and (ii) introduced an unconditional
volume fee. We ask: did this incentive change alter (a) the competitive
structure of the solver market and (b) the execution quality users
receive?

\textbf{Findings.} The reform reallocated trading \emph{value} by order
size. The robust, single-solver-proof signature is a monotone size
gradient: concentration fell in small orders and rose in large ones,
monotonically across four order-value buckets (Spearman $\rho$=1.00,
exact permutation p=0.042), with the largest effect in the 100k+ segment
--- precisely where protocol revenue, and thus the new revenue-linked
reward, is largest. Aggregate concentration also rose sharply at the
reform (HHI level break +0.055, top-share +0.082, the single largest
structural break of the 395-day sample landing 2--4 days after the
effective date), but a leave-one-solver-out test shows this aggregate
movement is substantially carried by the incumbent top solver; we
therefore lead with the order-size reallocation, which is robust to
dropping that solver. The active-solver count is unchanged, so the
reallocation operated through redistribution onto incumbents, not exit.
The timing is CoW-specific. We detect no change in average execution
quality, and a triple-difference around a February 2026 fee cut is
directionally consistent with the reward channel but underpowered (a
bounded null).

\textbf{Contribution.} We provide the first measurement of how a
solver-reward change reshapes competitive structure in an intent market,
paired with a mechanism --- a model and an order-size test --- that ties
the reallocation to the revenue-linked reward rather than to a single
firm's idiosyncratic rise. Methodologically, we combine an
interrupted-time-series design, changepoint detection, placebo-in-time
and permutation inference, a leave-one-solver-out test, a cross-venue
placebo, an order-size mechanism test, and a triple-difference on a
second governance event, all on public on-chain data. Substantively, a
reward change reshaped \emph{who captures value} --- shifting the
high-value segment toward inventory-rich incumbents --- without a
detectable change in the average price users paid.

We state our central claim precisely. It is \textbf{not} that
concentration rose only on CoW (it rose on the control venue too, later
and for separate reasons), nor that the \emph{number} of solvers winning
trades fell. It is that \textbf{CoW's trading-value concentration is
time-locked to CIP-74 and driven by the large-order segment the reform
most directly repriced}, whereas the control venue's increase is a
distinct, later, differently-driven event and the market-wide break is
months away.

\section{2. Institutional background}

\textbf{Solvers and rewards.} In CoW Protocol, orders are settled in
batches; for each batch a competitive auction selects a winning solver,
which executes the batch and earns a reward. The mechanism has evolved
through governance proposals (CIPs). The relevant changes:

\begin{itemize}
\tightlist
\item
  \textbf{CIP-67} ($\approx$20 May 2025): batch auction $\rightarrow$ fair combinatorial
  (multi-winner) auction.
\item
  \textbf{CIP-72} ($\approx$mid-Aug 2025): a quote earns a reward only if the
  solver later executes at least as well as it quoted.
\item
  \textbf{CIP-74} (effective $\approx$8 December 2025): fixed reward cap $\rightarrow$ cap =
  $\beta$$\cdot$(protocol revenue of the winning solution); plus an unconditional
  $\approx$2 bps volume fee. \textbf{Our primary treatment.}
\item
  \textbf{February 2026 amendment} (3 February 2026): volume fee on
  correlated/stable pairs cut from 2 bps to 0.3 bps --- a partial
  rollback, used as a second event.
\end{itemize}

\textbf{Why CIP-74 could change structure.} Tying the reward cap to
protocol revenue raises the payoff to winning high-revenue (large,
fee-bearing) solutions, plausibly advantaging solvers with the inventory
and capital to compete for large orders. We formalize this prediction.

\subsection{2.1 A solver-economics model}

Consider a unit order of value $v$ (USD). Solver $i$ can fill it at
private cost $c_i(v)$ --- routing, inventory, gas --- and competes by
offering the user execution surplus $s$. Under CIP-74 the winner earns a
reward capped at $\bar R = \beta\,\pi(v)$, where $\pi(v)=\phi v$ is the
protocol revenue, $\phi$ the ad-valorem fee rate, and $\beta\in(0,1]$ the
reward share. The fee $\phi v$ is deducted regardless of which solver
wins. Solver $i$'s payoff from winning at offered surplus $s$ is
\[
\small \pi_i(v,s) = \underbrace{\beta\phi v}_{\text{reward}} + \underbrace{m(v)\!-\!s}_{\text{margin}} - \underbrace{c_i(v)}_{\text{cost}} - \underbrace{\phi v}_{\text{fee}},
\]
where $m(v)$ is the gross spread. We model competition as Bertrand-style
in offered surplus: the winner is $\arg\max_i[m(v)-c_i(v)]$, the
lowest-cost solver, who captures the residual reward.

\textbf{Proposition 1 (the fee is competitively neutral).} Under Bertrand
competition in offered surplus, $\phi v$ enters every solver's payoff
identically and shifts all bids equally; it cannot change the identity of
the winner. An ad-valorem fee alone cannot reallocate wins --- it is a
transfer to the protocol, not a competitive lever. (Imperfect competition
changes the magnitude, not the direction.)

\textbf{Proposition 2 (the revenue-linked cap concentrates large
orders).} Pre-reform the cap is fixed, $\bar R_0$, independent of $v$;
post-reform it is $\beta\phi v$, increasing in order value. The marginal
return to winning rises with $v$ and is captured by the lowest-cost
solver. Inventory-rich solvers have $c_i(v)$ that grows slowly in $v$, so
their cost advantage is largest on large orders. The reform thus raises
the payoff to being the low-cost large-order solver, and value share on
large orders concentrates onto inventory-rich incumbents --- an effect
increasing in $v$, operating through \emph{value}, not count.

This yields three testable predictions: (i) concentration rises in value,
monotonically in order size, not in counts (\S6.4); (ii) cutting $\phi$ on
a subset of pairs shrinks their large-order reward pool and should
partially undo the concentration there (\S6.6); and (iii) this is the
causal counterpart to Chitra et al.~(2024). (The fee is not neutral on the
\emph{extensive} margin, where it changes which orders are worth solving;
conditional on an order being solved, Prop.~1 holds.)

\section{3. Related work and novelty}

Our paper sits at the intersection of three literatures and is, to our
knowledge, the first causal study of a solver-reward change.

\begin{itemize}
\tightlist
\item
  \textbf{Execution-quality measurement.} Bachu, Wan \& Moallemi (2024)
  formalize price improvement in order-flow auctions; Yuminaga et
  al.~(2025) measure execution welfare across CoW, UniswapX, and 1inch
  Fusion. Both are \emph{cross-sectional} and \emph{user-side}.
\item
  \textbf{Intent-market theory.} Chitra et al.~(2024) model solver entry
  and show restricted entry can be welfare-optimal. We provide the
  empirical, causal counterpart.
\item
  \textbf{Practitioner analysis.} CoW governance threads give
  \emph{descriptive} pre/post observations without control venues,
  identification, or inference. CIP-85 (``consistency rewards'') was
  later motivated partly by CIP-74's concentration effects --- independent
  corroboration of the phenomenon we measure.
\end{itemize}

More broadly, our mechanism connects to the DeFi-microstructure
literature on MEV and cross-chain execution (\"{O}z et al.~2025) and to the
dealer-inventory market-making literature (Barzykin, Bergault \& Gu\'{e}ant
2023), which explain why inventory-rich incumbents are best placed to
capture large-order flow. An adversarial novelty search found no prior
causal/event-study analysis of a CoW reward-rule change on concentration
or execution quality.

\section{4. Data}

All data are public and on-chain.

\begin{itemize}
\tightlist
\item
  \textbf{CoW solver concentration:} daily solver volume shares from
  \texttt{cow\_protocol\_ethereum.batches}, 1 March 2025 -- 30 March 2026
  (395 days): daily HHI of volume share, top-solver share, active-solver
  count.
\item
  \textbf{Order-size panel:} trade-level solver identity and USD value,
  bucketed \{0-1k, 1k-10k, 10k-100k, 100k+\}, for within-segment HHI.
\item
  \textbf{Pair-type panel:} solver volumes split by whether the pair is
  stable/correlated (both tokens in one of a stablecoin, ETH-correlated,
  or BTC-correlated set) vs.~non-stable, for the triple-difference (\S6.6).
\item
  \textbf{Control venue (UniswapX):} executor (\texttt{filler}) identity
  from decoded reactor \texttt{Fill} events; USD volume per fill from the
  swapper's outgoing transfer priced at the minute. Daily filler HHI
  computed identically.
\item
  \textbf{Execution quality:} trade-level price improvement vs.~a
  minute-level reference price, for CoW and UniswapX.
\end{itemize}

\emph{1inch Fusion was evaluated as a second control but its resolver
identity is not cleanly decodable on Dune, so a market-wide
DEX-aggregator HHI serves as the additional control.} We also compute
trade-count HHI to distinguish value from count concentration (\S6.4).

\section{5. Empirical strategy}

\textbf{Interrupted time series (ITS).} For each outcome $y_t$ we
estimate a level break at the CIP-74 effective date $t_0$:
\[\small y_t = \alpha + \tau t + \beta D_t + \gamma (t-t_0)D_t + \varepsilon_t,\quad D_t=\mathbf{1}[t\ge t_0],\]
with Newey--West (1987) HAC (14-lag) standard errors. $\beta$ is the
level break. Because this is a single autocorrelated series, we treat
the HAC asymptotic $p$-values as anti-conservative and rely on the
design-based inference below.

\textbf{Changepoint detection.} Independently of $t_0$, we locate the
single dominant mean-shift via a max-$F$ scan and ask how far it is from
CIP-74.

\textbf{Placebo-in-time.} We re-estimate the break at every interior
date and report where the CIP-74 break ranks.

\textbf{Cross-venue placebo.} We run the identical design on UniswapX,
whose rules did not change at $t_0$; identification rests on timing.

\textbf{Order-size mechanism and monotonicity test.} We estimate the ITS
break within each of four order-value buckets; Prop.~2 predicts the break
is monotone increasing in size. We test this with an exact permutation
test on the Spearman correlation between bucket and break (24
permutations), and re-run it dropping the largest solver.

\textbf{Leave-one-solver-out.} Because aggregate concentration is
mechanically sensitive to the largest solver, we recompute every headline
break with the top incumbent removed and decompose share changes.

\textbf{Triple-difference (second event).} The 3 February 2026 amendment
cut the volume fee on stable pairs (2$\rightarrow$0.3 bps), lowering the
fee-funded reward pool on those pairs only. Within a window held entirely
after CIP-74 we estimate
\[
\small
\begin{aligned}
y_{g,b,t} ={}& \alpha + \gamma_1 S_g + \gamma_2 P_t + \gamma_3 L_b \\
&+ \lambda_1 S_gP_t + \lambda_2 S_gL_b + \lambda_3 P_tL_b \\
&+ \delta\, S_gP_tL_b + \mu_t + \varepsilon,
\end{aligned}
\]
where $y$ is cell-level HHI; $S_g$, $P_t$, $L_b$ indicate stable pairs, the post-fee-cut period, and large (100k+) orders; $\mu_t$ are day fixed effects; and $\delta$
is the differential change in large-order concentration on stable pairs
after the cut. Prop.~2 predicts $\delta<0$. We \textbf{pre-specify the
100k+ bucket as the primary ``large'' definition} (the segment the model
singles out) and report 10k+ only as a sensitivity. Inference is by
randomization-in-time and placebo-in-unit, and we report the minimum
detectable effect so a null is read as bounded.

\textbf{Execution quality.} A venue $\times$ post DiD (Card \& Krueger
1994) and a synthetic control (Abadie, Diamond \& Hainmueller 2010) on
price improvement, with parallel-trends and randomization diagnostics.

\section{6. Results}

\begin{figure*}[t]\centering
\includegraphics[width=0.7\textwidth]{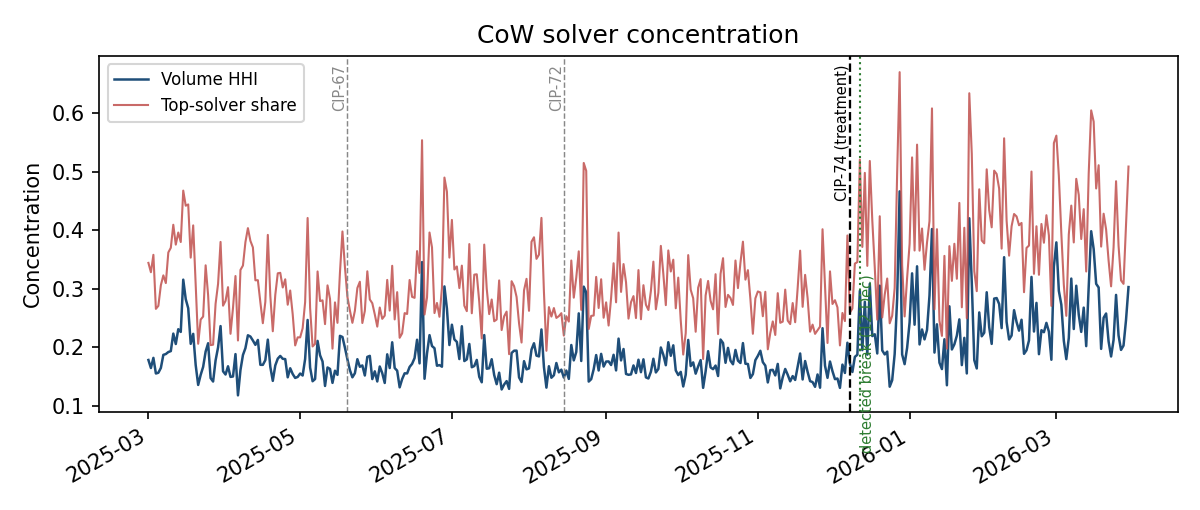}
\caption{CoW solver concentration (HHI and top-solver share), with CIP
event lines and the detected 12 December changepoint.}
\end{figure*}

\subsection{6.1 Trading value concentrated sharply at CIP-74}

Volume-weighted HHI rose from 0.176 to 0.241 (raw +0.065, +37\%;
trend-adjusted ITS level break +0.055) and top-solver share from 0.296 to
0.395 (raw +0.099, +34\%; level break +0.082). We foreground the
design-based inference over the HAC asymptotic $p$-values: the
placebo-in-time and permutation tests (\S6.2, \S6.4) are the credible
evidence, while the HAC values (HHI p=1.6$\times 10^{-8}$, top-share
p=8.2$\times 10^{-7}$) are reported for completeness only and are
anti-conservative on a single autocorrelated series. The break is
insensitive to the assumed date --- +0.055 at the 8 December effective
date, +0.063 at the 12 December detected changepoint (\S6.7). Active
solvers were unchanged ($\approx$21.5$\rightarrow$21.9, p=0.74) ---
\textbf{concentration rose by redistribution onto incumbents, not exit.}
The pre-CIP-74 trend was flat-to-declining (-0.008 per 100 days) and
reverses to +0.037 per 100 days after --- the trend \emph{changes sign} at
the reform. We show in \S6.7 that this \emph{aggregate} break is
substantially carried by the single largest solver; the load-bearing
result is the order-size gradient of \S6.4, which is not.

\subsection{6.2 The timing is precise}

The single dominant structural break of the entire 395-day sample lands
on 12 December (HHI) and 10 December (top-share) --- 2--4 days after the
8 December effective date. Placebo-in-time inference places the CIP-74
break near the most extreme of all interior dates: p$\approx$0.05 (HHI),
0.09 (top-share). Suggestive-to-significant on a single series,
corroborated by the changepoint location and the trend reversal.

\subsection{6.3 The timing is CoW-specific (cross-venue placebo)}

On UniswapX, concentration also rose in late December (HHI 0.365
$\rightarrow$ 0.522), but its dominant break is 31 December --- 23 days
after CIP-74 --- and placebo-in-time does not identify the CIP-74 date as
a break (p=0.33). We rule out a synchronized ``late-December wave'' three
ways. First, \emph{attribution}: CoW's December break is driven by a
single solver (Rizzolver, +0.080 share), while UniswapX's is driven by a
different executor (+0.183 share) --- different actors, 23 days apart.
Second, the \emph{cause} of the UniswapX break is idiosyncratic to that
venue (a single filler's onset). Third, \emph{market-wide context}: a
project-level DEX-aggregator HHI has its dominant break in May 2025 (220
days from CIP-74; placebo p=0.17 at the CIP-74 date). CoW (12 Dec),
UniswapX (31 Dec), and the market (May) are three distinct events with
distinct drivers.

\begin{figure*}[t]\centering
\includegraphics[width=0.7\textwidth]{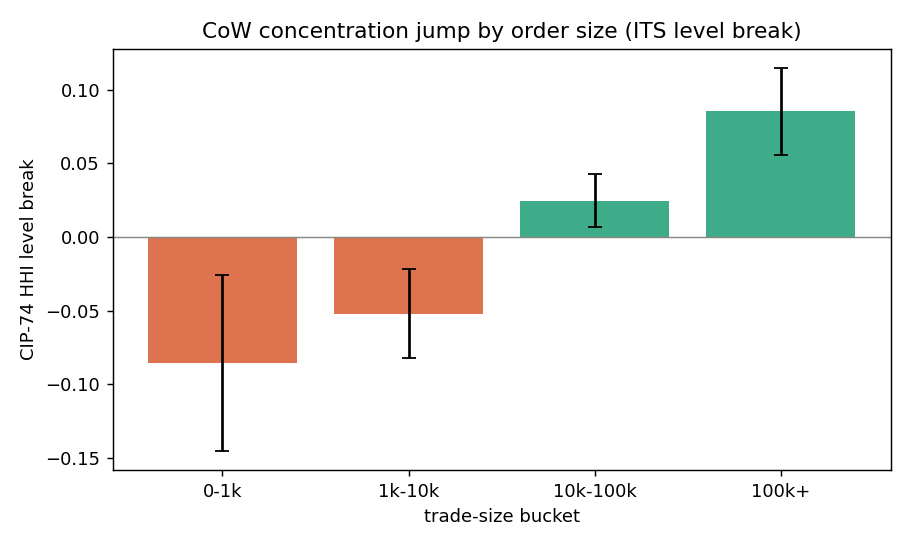}
\caption{ITS level break by order-size bucket: small orders
de-concentrate; 100k+ concentrates. The gradient is monotone (Spearman
$\rho$=1.00, exact permutation p=0.042) and survives dropping the largest
solver.}
\end{figure*}

\subsection{6.4 Mechanism: the effect is in large orders, and it is a value (not count) phenomenon}

The volume-HHI break is monotone in order size: 0-1k -0.086 (p=0.005),
1k-10k -0.052 (p=6.5$\times 10^{-4}$) --- small orders
\emph{de}concentrated --- while 10k-100k +0.025 (p=0.008) and \textbf{100k+
+0.085 (placebo-in-time p=0.016)}. The monotonicity is the sharpest
statement of the mechanism: the rank correlation between bucket and break
is perfect (Spearman $\rho$=1.00; exact permutation p=0.042), as Prop.~2
predicts. Critically, \textbf{the gradient survives dropping the single
largest solver} --- without Rizzolver the buckets remain monotone (-0.085,
-0.055, -0.000, +0.036; $\rho$=1.00, p=0.042) --- so the order-size
reallocation is not a one-firm artifact, even though the \emph{aggregate}
break is (\S6.7).

A trade-count decomposition disciplines the finding. By \emph{number of
trades}, concentration did not rise; it fell (count-HHI -0.060,
p=3.6$\times 10^{-4}$), and the large-order count-concentration is not
robust to placebo inference (100k+ count-HHI +0.052, placebo p=0.26). The
effect is therefore unambiguously a \emph{value} phenomenon: the incumbent
captured a larger share of trading \emph{value} via large orders, while
the \emph{number} of trades dispersed. This volume-up / count-down
divergence is exactly the signature the reform predicts. We verify it is
\emph{not} a whale-trade artifact: the incumbent's gain is backed by
$\approx$11,260 large trades post-reform ($\approx$100/day) at near-stable
mean size (\$316k$\rightarrow$\$435k), and persists for four months
(December 0.347, January--March 0.384); its \emph{rate} of large trades
actually fell (191$\rightarrow$100/day) while its \emph{share} rose --- a
sharper ``who wins'' restructuring than the whale alternative implies.

\subsection{6.5 No detectable execution-quality effect}

The naive venue $\times$ post DiD on price improvement is large (-25 bps)
but \textbf{not credible}: parallel pre-trends are rejected (F-test
p=2.8$\times 10^{-10}$) and randomization inference gives p=0.5. The
synthetic control yields a post-period gap of $\approx$+2.8 bps ---
statistically indistinguishable from zero. We report a \textbf{null} on
average execution quality. The null is informative: the design (N=333,264
trades, day-clustered) has a standard error of 2.5 bps, a minimum
detectable effect of $\approx$7.0 bps at 80\% power. We rule out average
execution-quality changes larger than $\approx$7 bps; the reform did not
move the typical price users received by an economically meaningful
margin, even as it concentrated value capture.

\subsection{6.6 The February fee cut: a triple-difference test (bounded null)}

The 3 February 2026 cut to the stable-pair fee (2$\rightarrow$0.3 bps)
lowered the fee-funded reward pool on those pairs, predicting (Prop.~2) a
partial reversal of large-order concentration there ($\delta<0$). On the
pre-specified 100k+ definition the triple-difference is in the predicted
direction but indistinguishable from zero: $\delta=-0.042$ (95\% CI
[-0.092, +0.008]; clustered-by-day p=0.099; randomization-in-time p=0.30;
placebo-in-unit p=0.66), within the minimum detectable effect of 0.072.
The sign is robust --- $\delta$ remains negative leave-Rizzolver-out
(-0.054) --- but the test cannot confirm the channel. We report a
\textbf{bounded null}: the fee cut was small and partial, so power is the
binding constraint, and the result is \emph{directionally consistent with}
the reward mechanism without confirming it.

On the cutoff we are explicit to avoid any appearance of specification
search. The 10k+ definition does yield a large, highly significant
clustered estimate ($\delta\approx-0.103$, p$\approx$2$\times 10^{-9}$),
but \textbf{this is not a reversal time-locked to 3 February}: its
randomization-in-time $p$-value is $\geq$0.26, so it reflects a
\emph{persistent} static stable$\times$large concentration gap --- a level
difference present throughout the post-CIP-74 window --- not an event
response. Under the design-matched inference (randomization-in-time) the
event test is null under \emph{both} cutoffs. Separately, aggregate
concentration did not fall after the cut, consistent with entrenchment,
but we do not lean on that descriptive observation given the
triple-difference is the identified test.

\subsection{6.7 Robustness}

\textbf{Leave-one-solver-out.} The aggregate breaks are substantially
carried by the incumbent top solver: removing Rizzolver, the aggregate
volume-HHI break falls from +0.055 to +0.004 and the top-share break flips
sign (+0.082 to -0.022). This is expected --- top-solver share mechanically
tracks the largest solver --- and is why we lead with the order-size
gradient (\S6.4), which is robust to the same exclusion. The reshuffle is
nonetheless multi-solver: Rizzolver (+0.126 mean share), Kipseli (+0.117),
and Tsolver (+0.084) gained while Barter (-0.193), Portus (-0.067), and
Quasi (-0.046) lost.

\textbf{Date sensitivity.} The aggregate break is insensitive to $t_0$:
HHI +0.055 at 8 December, +0.063 at the 12 December changepoint.

\textbf{Alternative measures.} The result does not depend on HHI: at $t_0$
the four-firm concentration ratio rises +0.059, the top-3 share +0.072,
and Shannon entropy falls -0.162 --- all consistent.

\begin{figure}[t]\centering
\includegraphics[width=\columnwidth]{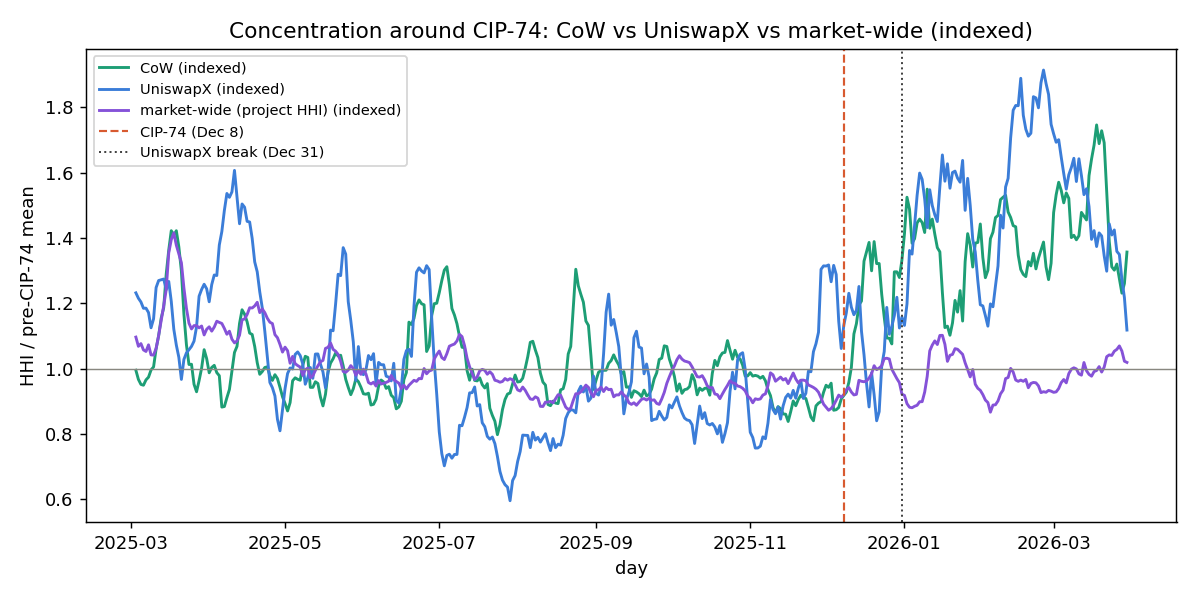}
\caption{Cross-venue context: CoW vs.~UniswapX vs.~market-wide HHI,
indexed to pre-CIP-74 = 1.0. Only CoW breaks at the reform date.}
\end{figure}

\section{7. Discussion}

The reform reshaped \emph{who wins} without a detectable change in
\emph{what users paid on average}. Tying solver rewards to protocol
revenue advantaged the large-order segment and shifted it toward
inventory-rich incumbents; the competitive margin moved, but average
execution quality did not visibly degrade. Incentive parameters that look
like accounting choices (how solvers are paid) are first-order
determinants of competitive structure, and structure can entrench quickly
and resist partial rollback --- consistent with the community's later
introduction of CIP-85 consistency rewards as a corrective.

\section{8. Limitations}

\begin{enumerate}
\def\labelenumi{\arabic{enumi}.}
\tightlist
\item
  \textbf{The aggregate break is substantially one solver; the order-size
  gradient is the robust result.} Leave-one-solver-out removes most of the
  aggregate HHI break and flips the top-share break (\S6.7). We rest the
  paper on the monotone gradient ($\rho$=1.00, p=0.042), which survives
  the exclusion, plus the trend reversal, changepoint timing, and
  cross-venue placebo.
\item
  \textbf{The reward channel is bounded, not confirmed.} The
  triple-difference is directionally consistent ($\delta$=-0.042) but a
  bounded null at this power (\S6.6); the mechanism rests on the model
  (\S2.1) and the order-size gradient, not the second event.
\item
  \textbf{One firm-level control venue} (UniswapX) plus a market-wide
  series; 1inch Fusion's executor identity is not cleanly decodable.
\item
  \textbf{The result is value, not trade-count, concentration} --- by count
  the market de-concentrated. It is not a claim about the number of
  solvers winning trades, and it is not a whale-trade artifact (\S6.4).
\item
  \textbf{Execution-quality null} is bounded by power ($\approx$7 bps), not
  a precise zero.
\item
  \textbf{Single chain} (Ethereum mainnet) and a single primary event;
  generalization is untested.
\end{enumerate}

\section{9. Conclusion}

Using a governance-dated solver-reward reform as a natural experiment, we
measure how incentive design reshapes the competitive structure of an
intent-based exchange. CoW Protocol's CIP-74 reallocated trading
\emph{value} by order size: large orders concentrated and small orders
dispersed, monotonically and robustly to dropping the largest solver, in
the segment the revenue-linked reward most directly repriced --- without
solver exit, without raising trade-count concentration, and without a
detectable change in average execution quality. A simple model explains
the pattern, and the order-size gradient is its empirical signature; a
triple-difference on a second fee event is directionally consistent but
underpowered. Solver-incentive design measurably reallocates who captures
intent-market value --- even when the average price users receive is
unchanged.

\begin{center}\rule{0.5\linewidth}{0.5pt}\end{center}

\section{Data and Code Availability}

All data are \textbf{public on-chain data}. The full replication package
--- code, SQL, and the frozen datasets (including the trade-level panels)
--- is committed to the anonymized repository\\
\url{https://github.com/ryonzhang/solver-reward-market-structure}
(tag \texttt{v1.0-preprint}); a fresh clone reproduces every table and
figure via \texttt{make all}. Data were pulled 2026-06-06 from Dune
Analytics and the CoW solver-competition API; the full source list and
schema notes are in the repository.

\begin{center}\rule{0.5\linewidth}{0.5pt}\end{center}

\section{References}

\begin{enumerate}
\def\labelenumi{\arabic{enumi}.}
\tightlist
\item
  B. Bachu, X. Wan, C. C. Moallemi. ``Quantifying Price Improvement in
  Order Flow Auctions.'' 2024. arXiv:2405.00537.
\item
  T. Chitra, K. Kulkarni, M. Pai, T. Diamandis. ``An Analysis of
  Intent-Based Markets.'' 2024. arXiv:2403.02525.
\item
  Y. Yuminaga, D. Chen, D. Sui. ``Execution Welfare Across Solver-based
  DEXes.'' 2025. arXiv:2503.00738.
\item
  B. \"{O}z, C. Ferreira Torres, C. Schlegel, B. Mazorra, J. Gebele, F.
  Rezabek, F. Matthes. ``Cross-Chain Arbitrage: The Next Frontier of MEV
  in Decentralized Finance.'' 2025. arXiv:2501.17335.
\item
  CoW DAO. CIP-67, CIP-72, CIP-74 (``Align solver rewards with protocol
  revenue and introduce a volume-based fee''), CIP-74 Retrospective, and
  CIP-85. CoW Protocol governance forum (forum.cow.fi) and docs.cow.fi,
  2025--2026.
\item
  A. Barzykin, P. Bergault, O. Gu\'{e}ant. ``Algorithmic market making in
  dealer markets with hedging and market impact.'' Mathematical Finance
  33(1):41--79, 2023. arXiv:2106.06974.
\item
  W. K. Newey, K. D. West. ``A Simple, Positive Semi-Definite,
  Heteroskedasticity and Autocorrelation Consistent Covariance Matrix.''
  Econometrica 55(3):703--708, 1987.
\item
  A. Abadie, A. Diamond, J. Hainmueller. ``Synthetic Control Methods for
  Comparative Case Studies.'' JASA 105(490):493--505, 2010.
\item
  D. Card, A. B. Krueger. ``Minimum Wages and Employment: A Case Study of
  the Fast-Food Industry in New Jersey and Pennsylvania.'' American
  Economic Review 84(4):772--793, 1994.
\end{enumerate}

\end{document}